\documentclass[twocolumn,twoside,slac_two]{revtex4}
\usepackage{graphicx}
\usepackage{fancyhdr}
\newcommand{\ifb}{\ensuremath{\mathrm{fb^{-1}}}}
\newcommand{\TeV}{\ensuremath{\mathrm{Te\kern -0.1em V}}}
\newcommand{\TeVc}{\ensuremath{\mathrm{Te\kern -0.1em V\!/}c}}
\newcommand{\TeVcc}{\ensuremath{\mathrm{Te\kern -0.1em V\!/}c^2}}
\newcommand{\GeV}{\ensuremath{\mathrm{Ge\kern -0.1em V}}}
\newcommand{\GeVc}{\ensuremath{\mathrm{Ge\kern -0.1em V\!/}c}}
\newcommand{\GeVcc}{\ensuremath{\mathrm{Ge\kern -0.1em V\!/}c^2}}
\newcommand{\MeV}{\ensuremath{\mathrm{Me\kern -0.1em V}}}
\newcommand{\MeVc}{\ensuremath{\mathrm{Me\kern -0.1em V\!/}c}}
\newcommand{\MeVcc}{\ensuremath{\mathrm{Me\kern -0.1em V\!/}c^2}}

\newcommand{\um}{\ensuremath{\mathrm{\mu m}}}

\newcommand{\babar}{\mbox{\slshape B\kern-0.1em{\small A}\kern-0.1em B\kern-0.1em{\small A\kern-0.2em R}}}

\def\mass{4143.0\pm2.9(\mathrm{stat})\pm1.2(\mathrm{syst})~\MeVcc }
\def\width{11.7^{+8.3}_{-5.0}(\mathrm{stat})\pm3.7(\mathrm{syst})~\MeVcc }
\def\yield{14\pm5 }
 
\def\massdifffit{1046.3\pm2.9~\MeVcc }
\def\widthfit{11.7^{+8.3}_{-5.0}~\MeVcc }

\begin{document}

%Title of paper
\title{Heavy quark meson spectroscopy at CDF}

\author{K. Yi}
\affiliation{Department of Physics and Astronomy, University of Iowa, Iowa City, IA 52242, USA}

\begin{abstract}
From a study of the $X(3872)$ mass and width based on the world's largest
sample of $X(3872)\rightarrow J/\psi\pi^+\pi^-$ decays, we find that our $X(3872)$ signal
is consistent with a single state, and leads to the most precise measurement 
of the $X(3872)$ mass.
We also report the recent evidence for a new narrow structure, 
$Y(4140)$, decaying to the $J/\psi \phi$ final state, in exclusive 
$B^+\rightarrow J/\psi\phi K^+$ decays in a data sample corresponding to an 
integrated luminosity of 2.7 \ifb collected at the CDF II detector. This 
narrow structure with its mass well above open charm pairs is unlikely  to 
be a candidate for a  
conventional charmonium state. 

\end{abstract}

\maketitle

\thispagestyle{fancy}

\section{Introduction}

It has been six years since the discovery of $X(3872)$~\cite{x3872discovery}, 
however, the nature of this state 
has not been clearly understood yet. 
Due to the proximity of the $X(3872)$ to the $D^0D^{*0}$  threshold, 
the $X(3872)$ has been proposed as a 
molecule composed of $D^0$ and $D^{*0}$. The $X(3872)$ has also been speculated 
to be two states nearby, as in some models like the $diquark-antidiqurk$ model. 
It is critical to make precise measurements of the mass and width of 
$X(3872)$ to understand its nature.
The  large  $X(3872)\rightarrow J/\psi\pi^+\pi^-$ sample accumulated at CDF enables us 
to  test the hypothesis of that the $X(3872)$ is composed of two states and to make a 
 precise mass measurement of $X(3872)$ if it is consistent with one state hypothesis.

There are many more states, similar to $X(3872)$, that have
charmonium-like decay modes 
but are difficult to place in the overall charmonium 
system~\cite{y3940,y4260discovery,y4320}.  These unexpected new 
states have introduced challenges to conventional $q\bar{q}$ meson model 
and revitalized interest in exotic mesons in the charm 
sector~\cite{conventional}, although the existence of exotic mesons
 has been discussed for many years~\cite{PDG}. 
Until recently all
of these new states involved 
only $c$ quark  and light quark ($u$, $d$) decay products.   
The $J/\psi\phi$ final state enables us to extend the exotic meson searches  
to $c$ quark and heavy $s$ quark decay products. 
An investigation of the $J/\psi\phi$ system produced in exclusive 
$B^+\rightarrow J/\psi\phi K^+$ decays with $J/\psi \rightarrow \mu^+ \mu^-$ 
and $\phi\rightarrow K^+K^-$ is reported in this note. 
Charge conjugate modes are included implicitly in this note.

\section{Measurement of the mass of $X(3872)$}

\begin{figure}[htb]
\centering
\includegraphics*[width=80mm]{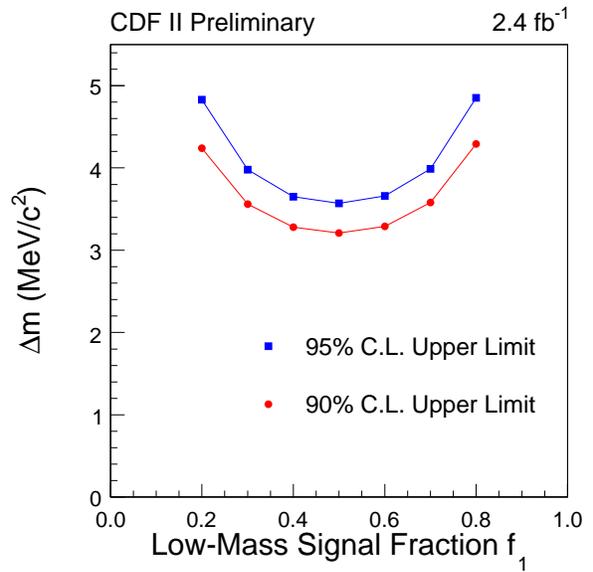}
\caption{
Dependence of the upper limit on the mass difference $\Delta m$ between two states on the 
low-mass state signal fraction $f_1$.
}
\label{Fig7}
\end{figure}

We tested the hypothesis 
of whether the observed X(3872) signal is composed of  two different states 
as predicted in some four-quark models  using the CDF inclusive $X(3872)$ sample. 
We fit the  $X(3872)$ mass signal with a Breit-Wigner function convoluted with a 
resolution function~\cite{x3872mass}. 
Both functions contain a width scale factor that is a free parameter in the fit 
and therefore sensitive to the shape of the mass signal. The measured width scale factor is 
compared to the values seen in simulations which assume two states with given mass 
difference and ratio of events. The resolution in the simulated events is corrected for the 
difference between data and simulation as measured from the $\psi(2S)$.
The result of this hypotheses test shows that the data is  consistent with  a single 
state. Under the assumption of two states with equal amount of observed events, we set a limit of
$\Delta m < 3.2 (3.6)$ \MeVcc at 90\% (95\%) C.L.
The limit for other ratios of events in the two peaks is shown in Fig.~\ref{Fig7}.

\begin{figure}[htb]
\centering
\includegraphics*[width=80mm]{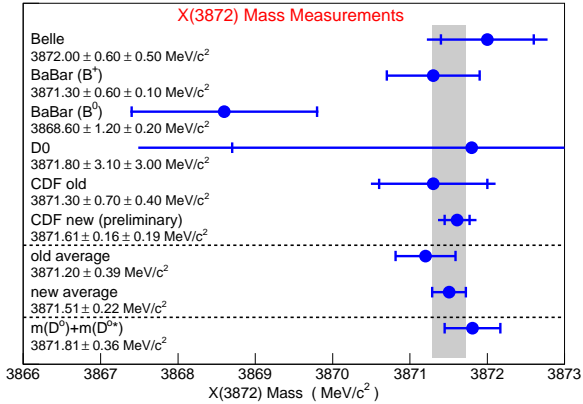}
\caption{
An overview of the measured $X(3872)$ masses from the experiments observing the $X(3872)$.
}
\label{Fig8}
\end{figure}

Since the $X(3872)$ is consistent with one peak in our test, 
we proceed to measure its mass in an 
unbinned maximum likelihood fit. The systematic uncertainties are determined from the difference 
between the measured $\psi(2S)$ mass and its world average value, the potential variation of the 
$\psi(2S)$ mass as a function of kinematic variables, and the difference in Q value 
between $X(3872)$  and $\psi(2S)$. We ignore systematics 
due to the fit model since they are negligible. The measured $X(3872)$ mass is:
$m(X(3872)) = 3871.61 \pm 0.16 (stat) \pm 0.19 (syst)$ \MeVcc, which  
is the most precise measurement to date as shown in Fig.~\ref{Fig8}~\cite{x3872mass,x3872massbelle}.

\section{Evidence for Y(4140) }

\begin{figure}[htb]
\centering
\includegraphics*[width=80mm]{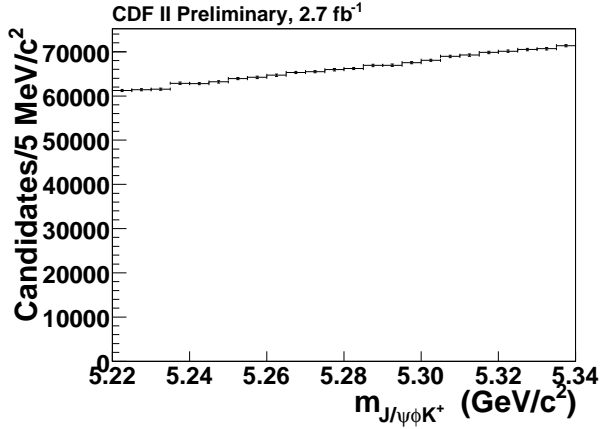}
\caption{The  $J/\psi\phi K^+$ mass before   
minimum ${L_{xy}(B^+)}$ and  kaon $LLR$ requirements.
}
\label{Fig1}
\end{figure}

We first reconstruct the $B^+\rightarrow J/\psi\phi K^+$ signal and 
then search for structures in the $J/\psi\phi$ mass spectrum~\cite{y4140}.
The $J/\psi\rightarrow \mu^+\mu^-$ events are recorded using a dedicated dimuon trigger. 
The  $B^+\rightarrow J/\psi\phi K^+$ candidates are reconstructed 
by combining a $J/\psi\to\mu^+\mu^-$ candidate, a $\phi\rightarrow K^+K^-$ candidate,  
and an additional charged track. 
Each track  is required to have at least 4  axial silicon  hits and have a transverse momentum 
greater than 400 \MeVc. 
The reconstructed mass of each vector meson candidate must lie within 
a suitable range from the nominal 
values ($\pm$50 \MeVcc~ for the $J/\psi$ and $\pm$7 \MeVcc~ for the $\phi$).  In the final $B^+$ 
reconstruction the $J/\psi$ is mass constrained, and the $B^+$ candidates must have
$p_T > 4$ \GeVc.
The $P(\chi^2)$ of the mass- and vertex-constrained fit to the 
$B^+\rightarrow J/\psi\phi K^+$ candidate is required to be greater than 1\%.

\begin{figure}[htb]
\centering
\includegraphics*[width=80mm]{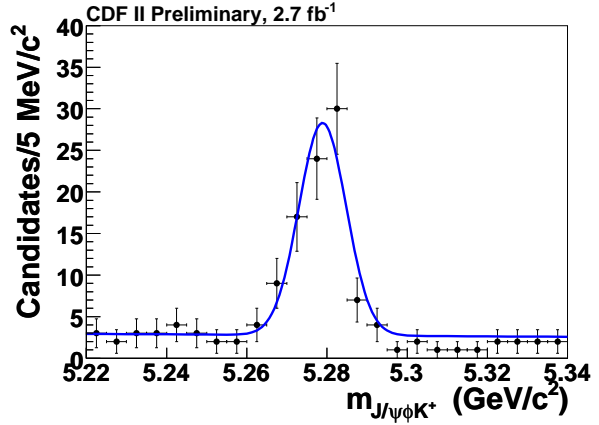}
\caption{
The $J/\psi\phi K^+$ mass after  
minimum ${L_{xy}(B^+)}$ and $LLR$ requirements; the solid line is a fit to 
the data with a Gaussian signal 
function and flat background function.
}
\label{Fig2}
\end{figure}

To suppress combinatorial background, we use ${dE/dx}$ and 
$\mathrm{TOF}$ information to identify all three kaons in the final state.
The information is summarized in a log-likelihood ratio ($LLR$), which 
reflects how well a candidate track can be positively identified 
as a kaon relative to other hadrons.
In addition, we require a minimum ${L_{xy}(B^+)}$ for the 
$B^+\rightarrow J/\psi\phi K^+$ candidate,
where ${L_{xy}(B^+)}$ is the projection onto ${\vec{p}_T(B^+)}$ 
of the vector connecting the primary vertex to the ${B^+}$ decay vertex.   
The  ${L_{xy}(B^+)}$ and   ${LLR}$ requirements for $B^+\rightarrow J/\psi\phi K^+$ 
are then chosen to 
maximize $\mathit{S/\sqrt{S+B}}$,  
where $\mathit{S}$ is the number of $B^+\rightarrow J/\psi\phi K^+$ signal events 
and $\mathit{B}$ is the number of background events  implied from the 
$B^+$ sideband.  
The requirements  obtained by  maximizing $\mathit{S/\sqrt{S+B}}$ are ${L_{xy}(B^+)}>500 ~\um$ 
and  ${LLR}>0.2$.

\begin{figure}[htb]
\centering
\includegraphics*[width=80mm]{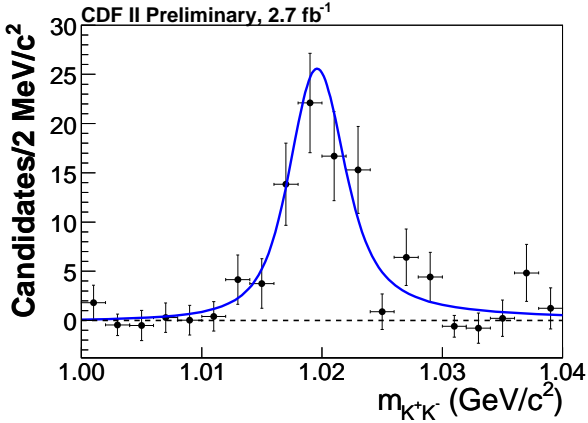}
\caption{
The $B^+$ sideband-subtracted   $K^+K^-$ mass without the $\phi$ mass window requirement.
The solid curve is a $P$-wave relativistic Breit-Wigner fit to the data. 
}
\label{Fig3}
\end{figure}

The invariant mass of $J/\psi\phi K^+$, after $J/\psi$ and $\phi$ mass window  
requirements, before and after the minimum ${L_{xy}(B^+)}$ and kaon ${LLR}$ requirements, 
are shown in Fig.~\ref{Fig1} and  Fig.~\ref{Fig2}, respectively. 
We do not see $B^+$ signal at 
all before the ${L_{xy}(B^+)}$ and kaon ${LLR}$ requirements, but we see clear 
$B^+$ signal after the  requirements. 
A fit with a Gaussian signal function and a 
flat background function  to the mass spectrum of $J/\psi\phi K^+$ (Fig.~\ref{Fig2}) 
returns a $B^+$ signal of $75\pm10(\mathrm{stat})$ events.
The ${L_{xy}(B^+)}$ and ${LLR}$  requirements reduce the background by a factor of 
approximately 20 000 
while keeping a signal efficiency of approximately 20\%.
We select  $B^+$ signal candidates with a mass 
within  3$\sigma$  of the nominal $B^+$ mass; 
the purity of the $B^+$ signal in that mass window is about 80\%.

\begin{figure}[htb]
\centering
\includegraphics*[width=80mm]{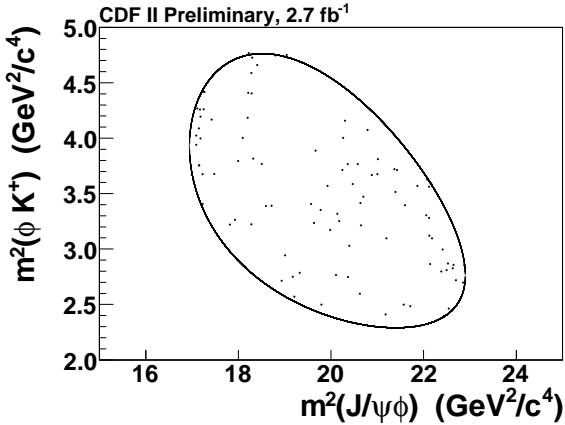}
\caption{
The Dalitz plot of ${m^2(\phi K^+)}$ 
versus ${m^2(J/\psi \phi)}$  in the  $B^+$ 
mass window. The boundary shows the kinematic allowed region.
}
\label{Fig4}
\end{figure}

The combinatorial background under the $B^+$ peak includes
$B$ hadron decays such as $B^0_s \rightarrow \psi(2S)\phi \rightarrow J/\psi \pi^+\pi^-\phi$,
in which the pions are misidentified as kaons.  However,
background events with misidentified kaons cannot
yield a Gaussian peak at the $B^+$ mass consistent with the
5.9 \MeVcc mass resolution.  
Figure~\ref{Fig3} shows the  $K^+ K^-$ mass  
from $\mu^+\mu^-K^+K^- K^+$ candidates within $\pm 3 \sigma$ of the nominal $B^+$ mass
with $B$ sidebands subtracted  before applying $\phi$ mass window requirement.
Using a smeared $P$-wave relativistic Breit-Wigner (BW)~\cite{pbw} line-shape fit to the spectrum 
returns a $\chi^2$ probability of 28\%.  
This shows that 
the $B^+ \rightarrow J/\psi K^+K^-K^+$ final state is well described 
by $J/\psi \phi K^+$.

\begin{figure}[htb]
\centering
\includegraphics*[width=80mm]{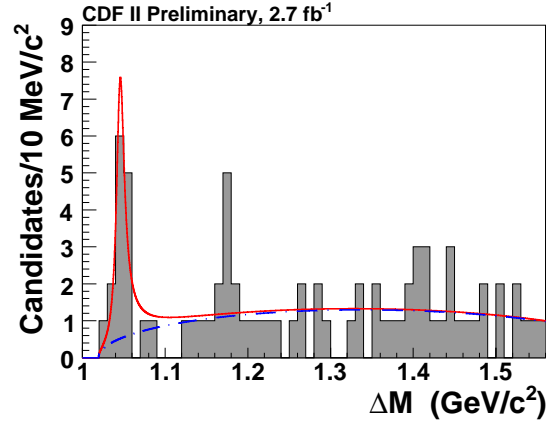}
\caption{
 The mass difference, $\Delta M$, between $\mu^+\mu^-K^+K^-$ and $\mu^+\mu^-$,  
 in the $B^+$ mass window. 
The dash-dotted curve is 
the background contribution and the red solid curve is the total unbinned fit.
}
\label{Fig5}
\end{figure}

We then examine the effects of detector acceptance and selection
  requirements using 
$B^+ \rightarrow J/\psi \phi K^+$ MC events simulated  by a phase space distribution. 
The MC events are smoothly distributed in the Dalitz plot and 
in the $J/\psi \phi$ mass spectrum. No artifacts were observed from MC events. 
Figure~\ref{Fig4} shows the Dalitz plot of ${m^2(\phi K^+)}$ 
versus ${m^2(J/\psi \phi)}$,   
and  Fig.~\ref{Fig5} shows 
the mass difference, $\Delta M= m(\mu^+\mu^-K^+K^-)- m(\mu^+\mu^-)$, for events in 
the $B^+$ mass window in our data sample. 
We examine the enhancement in the $\Delta M$ spectrum just above $J/\psi\phi$ threshold. 
We exclude the high--mass part of the spectrum beyond $1.56$ \GeVcc~ to avoid 
combinatorial backgrounds that would be expected from misidentified 
$B^0_s\rightarrow \psi(2S) \phi\rightarrow (J/\psi\pi^+\pi^-)\phi$ decays.
We model the enhancement by an $S$-wave relativistic BW 
function~\cite{sbw} convoluted with 
a Gaussian resolution function with the RMS fixed to 1.7 \MeVcc~obtained from MC,  
and use three--body phase space~\cite{PDG} to describe the background shape.
An unbinned likelihood fit to the $\Delta M$ distribution,   
as shown in Fig.~\ref{Fig5}, returns a yield of $\yield$ events,   
a $\Delta M$ of $\massdifffit$, and a width of $\widthfit$.
To investigate possible reflections, we examine the Dalitz plot and 
projections into  $\phi K^+$ and $J/\psi K^+$ spectrum.  
We find no evidence for any other structure in the $\phi K^+$ and $J/\psi K^+$ 
spectrum.

We use the log-likelihood ratio of $-2{\mathrm{ln}}(\mathcal{L}_0/\mathcal{L}_{{max}})$
to determine the significance of the enhancement, 
where $\mathcal{L}_0$ and $\mathcal{L}_{{max}}$ are the likelihood 
values for the null  hypothesis fit and signal hypothesis fit.
The $\sqrt{-2{\mathrm{ln}}(\mathcal{L}_0/\mathcal{L}_{{max}})}$ value 
is 5.3 for a pure three--body phase space 
background shape assumption. 
  We generate $\Delta M$ spectra using the background distribution
  alone, and search for the most significant fluctuation with 
$\sqrt{-2{\mathrm{ln}}(\mathcal{L}_0/\mathcal{L}_{{max}})}\ge 5.3$  in
  each spectrum in the mass range of 1.02 to 1.56 \GeVcc, with widths
  in the range of 1.7 (detector resolution) to 120 \MeVcc~(ten times the observed width).

\begin{figure}[htb]
\centering
\includegraphics*[width=80mm]{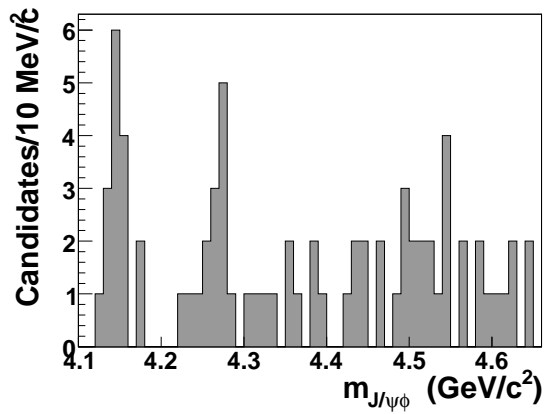}
\caption{
The $J/\psi\phi$ mass distribution  
 in the $B^+$ mass window.
}
\label{Fig6}
\end{figure}

The resulting $p$-value  from 3.1 million simulations 
is $9.3\times 10^{-6}$, corresponding to a significance of 4.3$\sigma$.
We repeat this process with a flat combinatorial non-B background and three--body PS 
for non-resonance $B$ background and we still get a significance of 3.8$\sigma$.

One's eye tends to be drawn to a second cluster of events
around 1.18 \GeVcc~in Fig.~\ref{Fig5}, or 
around 4.28 \GeVcc~in $J/\psi\phi$ mass as shown in Fig.~\ref{Fig6}.
This cluster is  close to one pion mass
above the peak at the $J/\psi\phi$ threshold. However, this cluster 
is statistically insufficient to infer the presence of a second structure.

\section{Summary}

Studies using CDF's $X(3872)$ sample, the largest in the world,
indicate that the $X(3872)$ is consistent with the 
one state hypothesis and this leads to the most precise mass measurement of $(X3872)$. 
The value is below, but within the uncertainties of, 
the $D^{*0}D^0$ threshold. The explanation 
of the X(3872) as a bound D*D system is therefore still an option.

The $B^+ \rightarrow J/\psi \phi K^+$   sample at CDF enables us 
to search for structure in the $J/\psi\phi$ mass spectrum,  
and we find  evidence for a narrow structure near the $J/\psi\phi$ 
threshold with a significance estimated to be at least  3.8$\sigma$.
Assuming an $S$-wave relativistic BW, the mass (adding $J/\psi$ mass) and width of this structure, 
including systematic  uncertainties,  are measured to be $\mass$ 
and $\width$, respectively.   This structure does not 
fit conventional expectations for a charmonium state 
because as a $c\bar{c}$ state it is expected to have a tiny branching ratio 
to $J/\psi\phi$ with its  mass well beyond open charm pairs.
We term the new structure the $Y(4140)$. The branching ratio  
of $B^+\rightarrow Y(4140) K^+, Y(4140)\rightarrow 
J/\psi\phi $ is estimated to be $9.0\pm3.4(stat)\pm2.9(B_{BF}))\times10^{-6}$.

\bigskip % extra skip inserted

\end{document}